\documentclass[12pt]{article}
\usepackage[a4paper,margin=2.5cm]{geometry}
\usepackage{amsmath,amssymb,graphicx,booktabs,mathptmx,xcolor}
\usepackage[T1]{fontenc}
\usepackage[labelfont=bf,font=small]{caption}
\usepackage{setspace}
\usepackage[round,authoryear,sort]{natbib}
\setcitestyle{aysep={},yysep={,},citesep={,}}
\usepackage[hidelinks]{hyperref}
\graphicspath{{figures/}}
\newcommand{\kpick}{\kappa_{\rm pick}}
\newcommand{\kobs}{\kappa_{\rm obs}}

\begin{document}
\begin{center}
{\Large\bfseries Fisher-information limits of Compton polarimetry for entanglement-based positronium imaging with pixelated CZT detectors\par}
\vspace{12pt}
{\large Haitao~Zhou$^{1,2,3,*}$, Ruonan~Wang$^{1,2,3}$, Jie~An$^{1,2,3}$, Min~Li$^{1,2,3}$, Junping~Lv$^{1,2,3}$, Yaqiong~Zhang$^{1,2,3}$, Hongliang~Wang$^{1,2,3}$, Zhifang~Wu$^{1,2,3}$ and Sijin~Li$^{1,2,3,*}$\par}
\vspace{8pt}
{\small
$^{1}$ Department of Nuclear Medicine, First Hospital of Shanxi Medical University, Taiyuan 030001, China\\
$^{2}$ Collaborative Innovation Center for Molecular Imaging of Precision Medicine, Shanxi Medical University, Taiyuan 030001, China\\
$^{3}$ CAEA Center of Excellence on Nuclear Technology Applications for Diagnosis, Treatment \& Transformation of Nuclear Medicine, Taiyuan 030001, China\\
$^{*}$ Authors to whom any correspondence should be addressed.\\
E-mail: \href{mailto:zht@sxmu.edu.cn}{zht@sxmu.edu.cn} and \href{mailto:sijinl@sxmu.edu.cn}{sijinl@sxmu.edu.cn}\par}
\end{center}
\section*{Abstract}
\noindent\textbf{Objective.} The polarisation correlation of annihilation photons, measured by Compton polarimetry, has been proposed as a PET biomarker of tissue microstructure. We quantify how many events this measurement needs, what limits that number and whether a small-animal CZT system reaches tissue-level contrasts.

\noindent\textbf{Approach.} We parameterised the measurement by the correlation coefficient $\kappa$ ($\kappa=1$ maximally entangled, $1/2$ separable), derived the classical Fisher information of double-Compton events and compared it with the quantum Fisher information. We related $\kappa$ to the annihilation channels of positronium in tissue and simulated pixelated CdZnTe (CZT) rings with Doppler broadening, cluster ordering, sequence reconstruction and in-body scattering. The model was validated against three published measurements and an independent GATE/Geant4 simulation.

\noindent\textbf{Main results.} A double-Compton event carries at most 0.12--0.14 units of Fisher information about $\kappa$ and 0.015 on average, 0.1--0.6\% of the quantum limit. Event-wise maximum likelihood reached the Cram\'er--Rao bound; the conventional ratio $R$ needed 1.2--2.3 times as many events. The model reproduced published modulation ratios of CZT (1.91 vs 1.85$\pm$0.04) and scintillator (2.46 vs 2.44$\pm$0.02) polarimeters, and GATE reproduced $\kappa=1$ and $1/2$ within 4\%. The observed coefficient is $\kobs=1-\alpha(1-\kpick)$ with pick-off share $\alpha=0.13$--0.29; it is sensitive to lipid fraction but not to oxygen, and published porous-polymer data give $\kpick=0.34$--0.58. In a 15-mm CZT module, ordering ambiguity, position errors and electron transport halved the information per event, and sequence reconstruction recovered it. With $^{18}$F and an eight-module ring, $\kappa$ was measured to $\pm0.01$ in a 10-mm tumour with half of its decays, whereas separating tissues by $\kappa$ alone required 6--74 times more decays than available. CZT energy resolution reduced in-body pre-scattering to 0.2--0.4\%.

\noindent\textbf{Significance.} These results set the statistical, estimator and detector requirements for entanglement-based positronium imaging, and show that the biomarker value of $\kappa$ is set by the polarisation state of pick-off photons.

\vspace{6pt}\noindent\textbf{Keywords:} PET, quantum-entangled PET, Compton polarimetry, Fisher information, positronium imaging, CZT detector, annihilation photons

\clearpage
\section{Introduction}

Positronium imaging extends positron emission tomography (PET) from where a tracer accumulates to the environment in which positrons annihilate \citep{Moskal2021,Bass2023,Moskal2024}. Three observables are under study: the mean ortho-positronium (o-Ps) lifetime, the ratio of three- to two-photon annihilations \citep{Fujimoto2026} and the polarisation correlation of the two annihilation photons \citep{Watts2021,Moskal2025,Moskal2026}. The last, measured by the azimuthal correlation of two Compton scatterings \citep{PryceWard1947,Snyder1948}, has been interpreted as a degree of quantum entanglement. In a porous polymer the measured correlation was smaller than expected for the maximally entangled state but larger than for separable photons, which was explained by assuming that photons from o-Ps pick-off annihilation are not entangled \citep{Moskal2025}. Because the pick-off fraction depends on free-volume size and composition, the correlation has been proposed as a biomarker of tissue type, and the first in vivo images of this quantity in a human subject have been reported \citep{Moskal2026FIH}; a roadmap towards quantum-entanglement PET has been outlined \citep{MoskalRoadmap2026}.

Three questions remain open. The first is what the measurement determines. A Compton polarimeter is insensitive to circular polarisation, and separable states can reproduce the double-Compton cross section of the entangled state \citep{Hiesmayr2019,Clarke2026}. Single-scattering observables built on the Pryce--Ward cross section therefore admit local hidden-variable descriptions and cannot certify entanglement on their own \citep{Clarke2026}; lower bounds on Bell-state fidelities that can be extracted from coarse-grained coincidences were derived only recently \citep{Sutter2026}, and it has been argued that the full double-Compton distribution of the maximally entangled state is reproduced only by that state \citep{Joos2026}. The second concerns what happens to the correlation when a photon scatters before detection: experiments disagree \citep{Watts2021,Ivashkin2023,Bordes2024,Tkachev2025}, and theoretical treatments of the evolution of the two-photon state are still being developed \citep{Caradonna2024,Caradonna2025,AskariPour2026}. The third concerns the pick-off channel: the interpretation of the J-PET result assumes that pick-off photons are separable \citep{Moskal2025,Moskal2026FIH}, an assumption that has not been derived from the quantum electrodynamics of annihilation with bound molecular electrons.

Underlying all three is a practical question: how many events are needed. In X-ray astronomy the statistics of single-photon polarimetry are well developed, through the minimum detectable polarisation and per-event Stokes-parameter estimators \citep{Weisskopf2010,Kislat2015}. For two-photon correlations the corresponding analysis has not been done. The most recent proposal for a Bell test with annihilation photons gives an order-of-magnitude rate estimate and defers the required number of events to a dedicated feasibility study \citep{Clarke2026}. Counts required to discriminate between two prescribed modulation levels, with finite acceptance, resolution and uniform background, have been derived within an idealised framework for coincidence-conditioned single-arm modulation \citep{Caradonna2026}. An estimation-theoretic treatment of the correlation coefficient itself, referred to the quantum limit and to realistic detectors, has not been given. For imaging the question is decisive, because tissue contrasts in the correlation coefficient are of order $10^{-2}$.

In a related study that is under review \citep{ZhouPMB}, we combined positronium lifetime, the three-to-two-photon ratio and the polarisation correlation in a source model and found that lifetime plus correlation is the combination that separates microstructure from oxygenation. That work assumed a constant analysing power and a usable fraction of double-Compton events. Here we derive these quantities from first principles and detector simulation and ask: (i) how much information a double-Compton event carries about the correlation coefficient, compared with the quantum limit; (ii) which estimator attains it; (iii) how the observed coefficient depends on annihilation channels, microstructure and oxygen, and on the pick-off state; (iv) how many usable events a pixelated CZT system delivers per decay; (v) whether in-body scattering biases the measurement; and (vi) whether tissue-level contrasts are reachable in a small-animal setting with $^{18}$F. The model is validated against three published measurements and an independent Geant4 simulation before it is used for these predictions.

\section{Methods}

\subsection{Measurement model}\label{sec:model}

We consider two back-to-back 511-keV photons, each undergoing a Compton scattering through polar angle $\theta_i$ and azimuth $\varphi_i$, with $\Delta\varphi$ the angle between the two scattering planes. A single Compton scattering is described by a positive-operator-valued measure parameterised by the analysing power \citep{Clarke2026}
\begin{equation}
\beta(\theta)=\frac{\sin^2\theta}{\epsilon+\epsilon^{-1}-\sin^2\theta},\qquad \epsilon=\frac{E'}{E}=\frac{1}{1+(E/m_ec^2)(1-\cos\theta)},
\label{eq:beta}
\end{equation}
which reaches its maximum at $\theta_{\rm opt}\approx 82^\circ$ for $E=m_ec^2$. Let $F(\theta)=\epsilon^2(\epsilon+\epsilon^{-1}-\sin^2\theta)$ be the unpolarised Klein--Nishina cross section (constants omitted). For a source without net polarisation, invariant under rotation about the line of flight and under parity, the joint distribution is
\begin{equation}
p(\theta_1,\theta_2,\Delta\varphi)\propto F(\theta_1)F(\theta_2)\left[1-\kappa\,\beta(\theta_1)\beta(\theta_2)\cos 2\Delta\varphi\right],
\label{eq:joint}
\end{equation}
which defines the linear-polarisation correlation coefficient $\kappa$. For the maximally entangled annihilation state $\kappa=1$; for a separable mixture of orthogonally polarised photons with uniformly distributed orientation $\kappa=1/2$; for uncorrelated photons $\kappa=0$. The widely used ratio $R=p(\Delta\varphi=90^\circ)/p(\Delta\varphi=0^\circ)=(1+a)/(1-a)$ with $a=\kappa\beta_1\beta_2$ follows directly. A uniform background fraction $b$ in $\Delta\varphi$ multiplies $\beta_1\beta_2$ by $1-b$. Because the marginal distribution of $(\theta_1,\theta_2)$ in (\ref{eq:joint}) does not depend on $\kappa$, all information on $\kappa$ resides in $\Delta\varphi$.

Equation~(\ref{eq:joint}) is the general form under the stated symmetries: the measurement determines one number, the common linear-polarisation correlation. Appendix~A shows that, for the Bell-diagonal states compatible with these symmetries, $\kappa$ constrains the concurrence $C$ only to the interval $[\max(0,2\kappa-1),\kappa]$, because the circular correlation is not measured. Throughout this paper we therefore treat $\kappa$ as the measured quantity and refer to entanglement only where this interval is invoked.

\subsection{Fisher information and quantum limit}\label{sec:fisher}

For fixed $\theta_1,\theta_2$, the conditional density of $\Delta\varphi$ is $(1-a\cos2\Delta\varphi)/2\pi$ with $a=\kappa\beta_1\beta_2$. Its Fisher information about $\kappa$ is (Appendix~B)
\begin{equation}
I_\kappa(\theta_1,\theta_2)=(\beta_1\beta_2)^2\,\frac{1}{a^2}\left(\frac{1}{\sqrt{1-a^2}}-1\right)\;\xrightarrow{a\to0}\;\frac{(\beta_1\beta_2)^2}{2}.
\label{eq:cfi}
\end{equation}
The information per double-Compton event is the average of (\ref{eq:cfi}) over the scattering-angle distribution $F(\theta_1)\sin\theta_1\,F(\theta_2)\sin\theta_2$, optionally restricted to an angular window. With $N$ independent events, the Cram\'er--Rao bound is $\sigma(\kappa)\geq(N\bar I_\kappa)^{-1/2}$.

To assess how much of the available information Compton polarimetry extracts, we computed the quantum Fisher information (QFI) for two one-parameter families of Bell-diagonal states. In family A, the maximally entangled state is mixed with the separable state described above, so that $\kappa\in[1/2,1]$; in family B, it is mixed with white noise, so that $\kappa\in[0,1]$. In both families the eigenbasis does not depend on the parameter, and the QFI equals the classical Fisher information of the Bell-basis weights:
\begin{equation}
F_Q^{(A)}=\frac{1}{\kappa(1-\kappa)},\qquad F_Q^{(B)}=\frac{9}{4(1+3\kappa)}+\frac{3}{4(1-\kappa)}.
\label{eq:qfi}
\end{equation}
As an intermediate reference we also evaluated (\ref{eq:cfi}) with $\beta_1=\beta_2=1$, which corresponds to an ideal linear-polarisation measurement with a continuous azimuth.

\subsection{Estimators and toy Monte Carlo}\label{sec:est}

We compared three estimators: event-wise maximum likelihood (ML) based on (\ref{eq:joint}); a moment estimator weighted by the analysing power, $\hat\kappa=-2\sum_j\beta_{1j}\beta_{2j}\cos2\Delta\varphi_j/\sum_j(\beta_{1j}\beta_{2j})^2$, which is the two-photon analogue of the per-event Stokes estimators used in X-ray polarimetry \citep{Kislat2015}; and the ratio $R$ formed from counts within $\pm22.5^\circ$ of $\Delta\varphi=0^\circ$ and $90^\circ$ and converted to $\kappa$ with the mean $\beta_1\beta_2$ of the accepted events. Events were drawn from (\ref{eq:joint}) for $\kappa=0.5$, 0.75 and 0.95, with $2\times10^4$ events per experiment and 300--400 repetitions, using scattering-angle windows of 60--110$^\circ$ and 20--160$^\circ$.

\subsection{Annihilation channels and tissue model}\label{sec:channels}

Photons from direct annihilation, para-positronium (p-Ps) self-annihilation and O$_2$-induced conversion of o-Ps to p-Ps were taken to be maximally entangled ($\kappa=1$). Pick-off photons were assigned an unknown coefficient $\kpick$. The observed coefficient of the two-photon sample is then
\begin{equation}
\kobs=1-\alpha\,(1-\kpick),
\label{eq:kobs}
\end{equation}
where $\alpha$ is the pick-off share of all two-photon annihilations. This is identical to the entanglement observable of \citet{ZhouPMB}, in which the partial-entanglement parameter $\eta$ corresponds to $\kpick=(1+\eta)/2$. We evaluated three hypotheses: H0, $\kpick=1/2$ (the working assumption of \citet{Moskal2025}); H$\eta$, $\kpick=3/4$; and H1, $\kpick=1$.

The channel shares were computed with the source model of \citet{ZhouPMB}, an absorbing Markov chain in which o-Ps and p-Ps interconvert through O$_2$ until annihilation, with parameters derived from published tissue positron-lifetime data \citep{Moskal2021,Moskal2023,Avachat2024,Stepanov2020,Moskal2026}. Microstructure was varied through a shift of the Tao--Eldrup free-volume radius \citep{Tao1972,Eldrup1981} and through the lipid fraction, at pO$_2$ = 30 mmHg. To test how the pick-off hypothesis propagates to oxygen estimation, we recomputed the lower bounds (floors) on $\sigma({\rm pO}_2)$ when four microstructure nuisance parameters are free within priors set by the spread between measured tissues \citep{ZhouPMB}.

\subsection{CZT detector Monte Carlo}\label{sec:czt}

We modelled a commercial pixelated CZT Compton-camera module (DT-PA05): CdZnTe (Cd$_{0.9}$Zn$_{0.1}$Te, 5.78 g cm$^{-3}$), $44\times44\times15$ mm$^3$, $22\times22$ pixels of 2 mm, sub-pixel lateral resolution 0.5 mm and depth resolution 1 mm (both treated as FWHM), and 1\% energy resolution at 662 keV. Compton attenuation was computed from the Klein--Nishina cross section with $Z/A=0.4173$; photoabsorption was parameterised as $\tau/\rho=0.0136\,(511\,{\rm keV}/E)^{2.8}$ cm$^2$ g$^{-1}$, which gives a total $\mu/\rho=0.0856$ cm$^2$ g$^{-1}$ at 511 keV. Coherent scattering, fluorescence escape and charge sharing were neglected. Transport within the crystal was unpolarised; the effect of the polarisation correlation on escape probabilities is of second order.

A photon was counted as a usable polarimetric event if (i) its full energy was deposited, (ii) its first interaction was Compton scattering, (iii) its interactions formed exactly two pixel clusters, the first containing only the first interaction, and (iv) both clusters exceeded an energy threshold $E_{\rm th}$ (10, 15 or 30 keV). The scattering angle was reconstructed from the energy of the first cluster, and the azimuthal resolution was propagated as a modulation factor $\exp(-2\sigma_\varphi^2)$ with $\sigma_\varphi=\sqrt2\,\sigma_\perp/L_\perp$, where $L_\perp$ is the transverse cluster separation. Events with three or more clusters, which require sequence reconstruction, were counted separately as an upper bound. This idealised model, which assumes correct cluster ordering, was used for the ring geometry and the pre-scattering study.

We simulated a single module, then two rings with a point source at the centre: G8, eight modules in an octagon with crystal front faces at $R=10.5$ cm, which reflects the 83.5-mm housing; and G16, sixteen crystals of the same size forming a gapless polygon ($R=11.06$ cm), which represents tighter packing. Photons could cross between modules. The per-decay Fisher information was the probability that both photons gave usable events multiplied by the mean of (\ref{eq:cfi}) over these pairs, with $b=0.05$ and a time-averaged live fraction of 0.596 for double coincidences at 10 MBq, as in \citet{ZhouPMB}. Random coincidences between usable single events in opposite modules were estimated for a 24-ns window at 10 MBq.

\subsection{Detector realism}\label{sec:real}

A second single-module simulation removed the main idealisations. (i) Bound-electron Doppler broadening was included in the impulse approximation: the scattered energy was shifted by $\delta E'=p_z|\mathbf q|/[137\,(1+E(1-\cos\theta)/m_ec^2)]$, with $|\mathbf q|$ the momentum transfer and the electron momentum projection $p_z$ drawn from a Gaussian of standard deviation 1.5 atomic units (1 and 2 a.u.\ as variants). (ii) Positions were the energy-weighted centroids of the interactions within each 2-mm pixel, smeared by the sub-pixel lateral (0.5 mm FWHM) and depth (1 mm FWHM) resolutions; energies were smeared by the energy resolution and a full-energy window of $\pm2\sigma$ was applied. (iii) The order of two clusters was not assumed. Two rules were compared: the larger deposit is the first scattering \citep{Watts2021}, and, using the incident direction known from the line of response, the order whose geometric scattering angle agrees best with the angle computed from the energies. (iv) Events with three or four clusters were reconstructed by testing all orderings and choosing the one that minimised the summed squared difference between geometric and kinematic scattering cosines along the sequence.

For each accepted event we recorded the true analysing power $\beta(\theta_{\rm true})$ of the first scattering, the reconstructed angle $\theta_{\rm rec}$ and the error $\varepsilon$ of the reconstructed scattering-plane axis, which includes ordering errors. The modulation retained by reconstructed events is $\beta\cos2\varepsilon$. Because a real analysis can use only $\theta_{\rm rec}$, the effective analysing power was calibrated in 18 bins of $\theta_{\rm rec}$, $\beta_{\rm eff}(\theta_{\rm rec})=\langle\beta(\theta_{\rm true})\cos2\varepsilon\rangle$, as is done with a calibration source in an experiment; the information per pair was then the average of (\ref{eq:cfi}) with $\beta_{\rm eff}$ over random pairs of accepted events ($\kappa=0.926$, $b=0.05$). The ratio of the information per pair of interacting photons to that of the idealised model, taken from the independent simulation of section~\ref{sec:gate}, was used to rescale the ring results. One parameter at a time was varied to obtain the sensitivity to the photoabsorption cross section ($\pm10\%$), pixel size (1 and 3 mm), crystal thickness (10 and 20 mm), threshold (10 and 30 keV), energy resolution (2\%) and Doppler width.

The total attenuation coefficient of the model was compared with the tabulated values for Cd$_{0.9}$Zn$_{0.1}$Te obtained from the elemental tables of \citet{Hubbell2004} by mass-weighted addition.

\subsection{Independent Monte Carlo}\label{sec:gate}

The detector model and the correlation model were checked with GATE 10 \citep{Sarrut2022}, which is based on Geant4 \citep{Allison2016}, using the Livermore polarised electromagnetic physics, which includes Doppler broadening, fluorescence and electron transport. Two simulations were run. In the first, 511-keV photons from a point 10.5 cm in front of a DT-PA05 module were tracked; the deposits were processed with the same clustering, ordering and calibration as above. In the second, $2\times10^6$ positrons annihilated at rest in a 1-mm water sphere between two $30\times30\times1.5$ cm$^3$ CZT slabs at $\pm10.5$ cm, with Geant4's entangled annihilation-photon option enabled or disabled; $\kappa$ was fitted by maximum likelihood with (\ref{eq:joint}) to the true first-Compton angles and scattering planes of both photons.

\subsection{Comparison with published measurements}\label{sec:valid}

Three published measurements were reproduced with the model. (i) \citet{Watts2021} measured the modulation ratio $R$ with two CZT polarimeters ($1\times1$ cm$^2$ detectors with sub-millimetre pixels, 4.35 cm from the source), selecting events with two pixels and assigning the first scattering to the larger deposit; they reported $R=1.85\pm0.04$ for scattering angles of 70--110$^\circ$ and $1.95\pm0.07$ for 93--103$^\circ$. We simulated their geometry with the transport model of section~\ref{sec:czt}, reconstructed angles and azimuths from pixel centres with their ordering rule, and computed $R$ for $\kappa=1$ and $1/2$. (ii) \citet{Ivashkin2023} used scintillator scatterers with 16 azimuthal counters of 22.5$^\circ$ and scattering angles of 80--100$^\circ$, reporting $R=2.435\pm0.018$; we folded (\ref{eq:joint}) with their angular acceptance. (iii) \citet{Moskal2025} measured $R$ for photons from the porous polymer XAD-4 at 82$^\circ$ and 94$^\circ$, with circular angular regions of 20$^\circ$ radius, and quoted ideal and detector-simulated ratios for the maximally entangled and separable cases. We used their simulated ratios as a scale, checked that our model reproduces their separable reference, and inverted their measured ratios to $\kobs$ and, through (\ref{eq:kobs}) with their pick-off share, to $\kpick$.

\subsection{In-body pre-scattering}\label{sec:body}

A photon that scatters by a small angle in the body may remain inside the photopeak window, and its correlation after scattering is disputed \citep{Watts2021,Ivashkin2023,Tkachev2025,Moskal2026FIH}. We therefore treated the post-scattering coefficient $\kappa_s$ as unknown in $[0,\kappa]$, so that the bias is bounded by $f_s\kappa$, where $f_s$ is the fraction of accepted pairs with at least one pre-scattered photon. The mouse was modelled as an infinite water cylinder of radius 1.0 or 1.5 cm with the source on the axis or 0.5--1.0 cm off axis. Only Compton scattering in water was tracked, and photons were accepted if they reached a module of G8. The photopeak window was $\pm2\sigma$ of the CZT energy resolution at 511 keV (FWHM 5.8 keV), which admits scattering angles below 8.0$^\circ$; a case without an energy window was simulated for comparison with detectors of low energy resolution. The quantity of interest is the difference in $f_s\kappa$ between source positions, since a depth-dependent bias could mimic a tissue contrast.

\subsection{Imaging scenarios}\label{sec:scen}

We considered $^{18}$F with 10 MBq injected in a mouse and acquisition until complete decay. Two regions were used: a 10-mm tumour (0.5 mL at 400 kBq mL$^{-1}$, $1.9\times10^9$ decays) and a liver of about 1.5 mL ($5.7\times10^9$ decays). The decays required per region were $N=1/(I_{\rm dec}\,\delta\kappa^2)$ for a target precision $\delta\kappa$, and $N=50/(I_{\rm dec}\,\Delta\kappa^2)$ for a 5$\sigma$ separation of two regions whose coefficients differ by $\Delta\kappa$.

\subsection{Implementation}

All calculations used Python with NumPy; the source model of \citet{ZhouPMB} additionally used SciPy, and the independent simulations used GATE 10.1. Analytical Fisher information was checked against direct numerical integration, and the QFI expressions against numerical differentiation of the Bell weights. Code and numerical results are archived at Zenodo (see Data availability).

\section{Results}

\subsection{Benchmarks}

The analysing power reached $\beta_{\max}=0.6918$ at $81.66^\circ$, in agreement with \citet{Clarke2026}. With both photons at 82$^\circ$, (\ref{eq:joint}) gave $R=2.835$ for $\kappa=1$ and $R=1.629$ for $\kappa=1/2$, reproducing the values usually quoted (2.85 and 1.63) \citep{Ivashkin2023,Moskal2025}. Equation~(\ref{eq:cfi}) agreed with numerical integration to six digits for $\kappa$ between 0 and 1, and (\ref{eq:qfi}) with numerical differentiation to five digits.

\subsection{Comparison with published measurements}

Table~\ref{tab:valid} summarises the comparison. For the CZT polarimeter of \citet{Watts2021}, the model predicted $R=1.91$ (70--110$^\circ$) and 1.98 (93--103$^\circ$) for $\kappa=1$, against measured values of $1.85\pm0.04$ and $1.95\pm0.07$; for $\kappa=1/2$ it predicted 1.37 and 1.40. The larger-deposit rule ordered 69\% of events correctly, close to the 60\% reported, and pixelisation and misordering lowered $R$ from 2.06 and 2.15. For the scintillator polarimeter of \citet{Ivashkin2023}, folding with the counter width gave $R=2.46$, against $2.435\pm0.018$. Both measurements agree with $\kappa=1$ for direct annihilation in these sources and validate our treatment of angular acceptance, pixelisation and misordering.

For the J-PET data \citep{Moskal2025}, our model reproduced the separable reference from their simulated ratios ($\kappa=0.508$ and 0.512 at 82$^\circ$ and 94$^\circ$). Their measured ratios correspond to $\kobs=0.796\pm0.016$ and $0.869\pm0.019$; with their pick-off share this gives $\kpick=0.34\pm0.05$ and $0.58\pm0.06$. Both values exclude $\kpick=1$ by at least seven standard deviations. The two angles differ from each other by three standard deviations, and the XAD-4 mixture implied by the J-PET simulation at 94$^\circ$ (0.895) differs from that implied by the stated pick-off share (0.845); fixing $\kpick$ more precisely requires the uncorrected counts, which have not been published.

\subsection{How much information a double-Compton event carries}

\begin{figure}[h]
\centering\includegraphics[width=\textwidth]{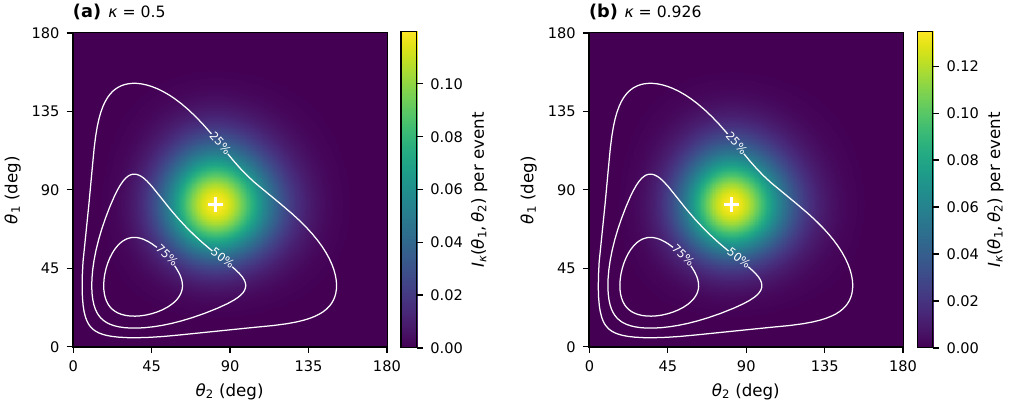}
\caption{Fisher information about $\kappa$ per double-Compton event as a function of the two scattering angles, for (a) $\kappa=0.5$ and (b) $\kappa=0.926$. White contours: relative density of double-Compton events (25, 50, 75\% of maximum). Cross: optimal angle 81.7$^\circ$.}\label{fig:map}
\end{figure}

Figure~\ref{fig:map} shows $I_\kappa$ over the $(\theta_1,\theta_2)$ plane. The information peaks where both photons scatter near 82$^\circ$, at 0.120 for $\kappa=0.5$ and 0.139 for $\kappa=1$, so that $\delta\kappa=0.01$ needs $7$--$8\times10^4$ such events. The Klein--Nishina distribution, however, places most scatterings at forward angles where $\beta$ is small. Averaged over all angles, a double-Compton event carries 0.0146--0.0157, and $\delta\kappa=0.01$ requires $6.4$--$6.9\times10^5$ events, about nine times more than the optimal-angle estimate.

\begin{figure}[h]
\centering\includegraphics[width=\textwidth]{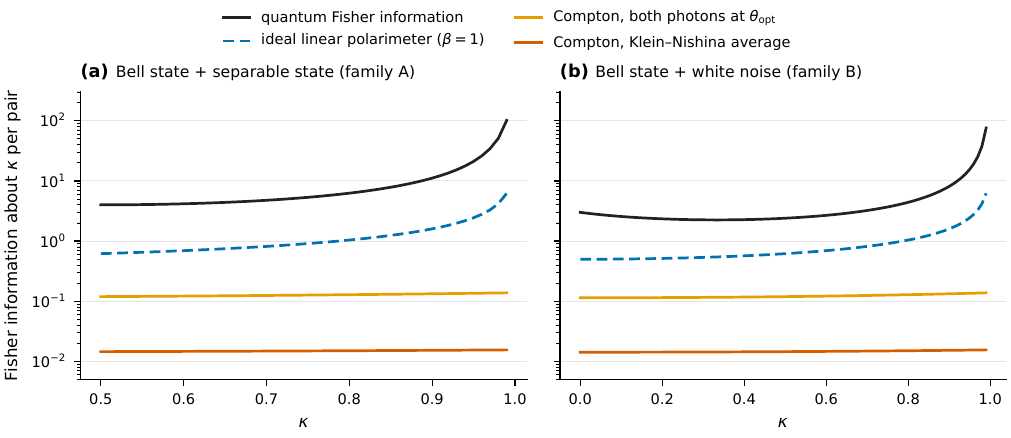}
\caption{Information about $\kappa$ per photon pair: quantum Fisher information, ideal linear-polarisation measurement, Compton polarimetry at the optimal angle and Compton polarimetry averaged over the Klein--Nishina distribution, for (a) family A and (b) family B.}\label{fig:chain}
\end{figure}

Figure~\ref{fig:chain} places these numbers on the scale of the quantum limit. For family A at $\kappa=0.5$ the QFI is 4.0; an ideal linear-polarisation measurement extracts 0.62 (15.5\%); Compton polarimetry at the optimal angles extracts 0.12 (3.0\%); and Klein--Nishina-averaged Compton polarimetry extracts 0.015 (0.37\%). For family B the corresponding fractions at $\kappa=0.5$ are 25.8\%, 5.0\% and 0.61\%. The first step reflects the blindness to circular polarisation and the restriction to a single azimuthal outcome; the second reflects $\beta<0.7$; the third reflects the scattering-angle distribution. As $\kappa\to1$ the QFI of family A diverges, whereas the Compton information stays finite, so the efficiency falls further (0.14\% at $\kappa=0.9$).

\subsection{Angular selection and estimators}

\begin{figure}[h]
\centering\includegraphics[width=\textwidth]{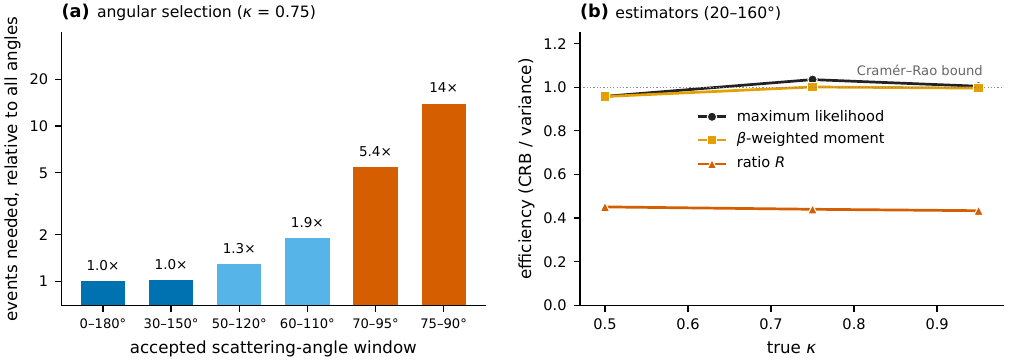}
\caption{(a) Double-Compton events needed for a given precision when a scattering-angle window is imposed, relative to using all angles ($\kappa=0.75$). (b) Efficiency of three estimators (Cram\'er--Rao bound divided by Monte Carlo variance) with a 20--160$^\circ$ window and $2\times10^4$ events.}\label{fig:est}
\end{figure}

Selecting a window of scattering angles raises the information per accepted event but discards events, and the product always decreased (figure~\ref{fig:est}(a)). Relative to using all angles, the windows 60--110$^\circ$, 70--95$^\circ$ and 75--90$^\circ$ required 1.9, 5.4 and 14 times as many double-Compton events. A window of 30--150$^\circ$ lost only 1\%.

Event-wise ML and the $\beta$-weighted moment estimator were unbiased and reached the Cram\'er--Rao bound within Monte Carlo error at all three values of $\kappa$ (efficiency 0.96--1.03; figure~\ref{fig:est}(b)). The ratio $R$ had an efficiency of 0.43--0.45 with the 20--160$^\circ$ window, that is it needed 2.2--2.3 times as many events. With the narrower 60--110$^\circ$ window the penalty was smaller (variance ratio 1.24) because $\beta$ varies less within the window.

\subsection{Annihilation channels, microstructure and the pick-off hypothesis}

Across the tissue parameter set, the pick-off share of two-photon annihilations $\alpha$ ranged from 0.13 (porcine liver) to 0.29 (human adipose; lifetime data of \citet{Moskal2023}), giving $\kobs$ between 0.854 and 0.934 under H0 (table~\ref{tab:tissue}). Under H$\eta$ the coefficients moved halfway towards 1, and under H1 they were all exactly 1.

\begin{figure}[h]
\centering\includegraphics[width=\textwidth]{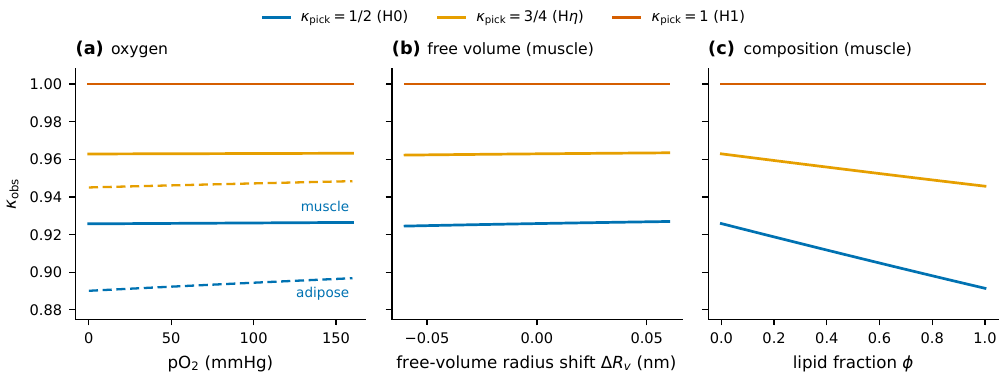}
\caption{Observed correlation coefficient as a function of (a) pO$_2$, (b) free-volume radius shift and (c) lipid fraction, for three pick-off hypotheses. Muscle base tissue in (b) and (c), pO$_2$ = 30 mmHg.}\label{fig:kobs}
\end{figure}

The coefficient was sensitive to composition but almost insensitive to oxygen (figure~\ref{fig:kobs}). In muscle under H0, raising pO$_2$ from 20 to 40 mmHg changed $\kobs$ by $1.0\times10^{-4}$, a free-volume radius shift of $\pm0.01$ nm by $4.0\times10^{-4}$, and a lipid fraction of 0.1 by $-3.6\times10^{-3}$. Contrasts between tissue pairs were $\Delta\kappa=0.0193$ (muscle vs porcine adipose), 0.0080 (muscle vs liver) and 0.0055 (cardiac myxoma vs adipose) under H0, and half of these under H$\eta$.

Table~\ref{tab:floor} shows how the hypothesis propagates to oxygen estimation. In the analysis of \citet{ZhouPMB}, adding the correlation to lifetime events lowers the floor on $\sigma({\rm pO}_2)$ from 2214 to 100 mmHg in muscle and from 339 to 14 mmHg in adipose. These floors were identical under H0 and H$\eta$, because the floor depends on how the observables vary with the parameters and not on the size of that variation; the number of decays needed to approach the floor scales as $(1-\kpick)^{-2}$. Under H1 the correlation coefficient carries no information and the floor reverts to that of lifetime alone.

\subsection{Usable events in CZT}

\begin{figure}[h]
\centering\includegraphics[width=\textwidth]{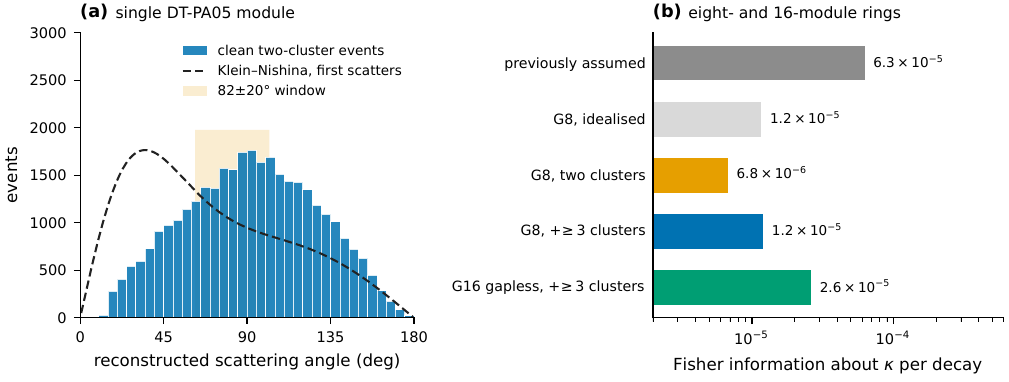}
\caption{(a) Reconstructed scattering angle of clean two-cluster, full-energy events in a DT-PA05 module, with the Klein--Nishina distribution of first scatterings (peak-scaled) and the 82$\pm$20$^\circ$ window. (b) Fisher information about $\kappa$ per decay (live time included) assumed by \citet{ZhouPMB} and obtained here with the idealised model and with the GATE-corrected response using two clusters only or including sequence reconstruction of events with three or four clusters.}\label{fig:czt}
\end{figure}

In a single 15-mm module (table~\ref{tab:czt}), 47.0\% of incident 511-keV photons interacted; of these, 50.4\% deposited their full energy and 84.1\% interacted first by Compton scattering. Only 17.3\% of interacting photons produced a clean two-cluster, full-energy event, and 6.6\% did so within the 82$\pm$20$^\circ$ window commonly used for $R$. The main loss is lateral escape: a photon scattered near 82$^\circ$ carries about 250 keV sideways and must be absorbed within the same 44-mm crystal. Thicknesses of 10 and 20 mm gave windowed fractions of 6.1\% and 6.8\%, and thresholds between 10 and 30 keV changed the fractions by less than 0.002. Clean events had an approximately symmetric distribution of reconstructed angles centred near 90$^\circ$ (figure~\ref{fig:czt}(a)), unlike the forward-peaked distribution of first scatterings. Within the window, the effective analysing power $\langle\beta_1\beta_2\rangle$ was 0.391, and the 2-mm pixels reduced the modulation by less than 1\% (median factor 0.994). A further 8.5\% of interacting photons produced three or more clusters with full energy.

In the eight-module ring, the probability that both photons interacted was $2.56\times10^{-2}$ per decay; usable pairs occurred at $8.7\times10^{-4}$ per decay over all angles and $1.3\times10^{-4}$ within the window. Only 1.7\% of usable events involved two modules, because the housings leave wide gaps between crystals. The Fisher information per decay including live time was $1.15\times10^{-5}$ with all angles and ML, and $5.8\times10^{-6}$ with the window (figure~\ref{fig:czt}(b)). In the gapless 16-module ring, both photons interacted with probability $4.6\times10^{-2}$, 5.1\% of usable events crossed modules, and the information per decay rose to $2.55\times10^{-5}$. Random coincidences contributed 0.5--0.6\% of usable pairs at 10 MBq in both geometries.

\citet{ZhouPMB} assumed a usable single-photon fraction of 0.22 (range 0.15--0.30) within the window. The simulation gives 0.066, which reduces the information per decay by a factor of 11 with windowed analysis and of 5.4 with all angles and ML. The per-event information assumed there (effective analysing power 0.41) was consistent with the Klein--Nishina average within the window (0.415).

The realistic model changed these numbers in two opposite directions (table~\ref{tab:real}). With positions from sub-pixel centroids and the kinematic ordering rule, 19\% of interacting photons gave usable two-cluster events, and the order was correct for 70\% of them. Misordering is intrinsic near 90$^\circ$, where both deposits are close to 256 keV, and it reverses the scattering-plane axis only partly; together with position errors it reduced the mean $\cos2\varepsilon$ to 0.84, and 14\% of events had axis errors above 20$^\circ$. After calibration in $\theta_{\rm rec}$, the information per usable pair was 0.013, against 0.023 in the idealised model, so that two-cluster events alone gave 0.68 times the idealised information per pair of interacting photons. Doppler broadening of 1--2 a.u.\ had no measurable effect (relative change below 3\%), because the angular resolution was dominated by the lever arm between clusters. Ordering by the larger deposit lowered the information by a further 19\%. Including events with three or four clusters raised the usable fraction to 27\%; their first pair was identified correctly in 87\% of cases, and the information per pair of interacting photons rose to 1.26 times the idealised two-cluster value.

Table~\ref{tab:design} lists the sensitivity. Crystal thickness had the largest effect (0.75 and 1.21 for 10 and 20 mm), followed by pixel size (1.12 and 0.87 for 1 and 3 mm) and threshold (1.03 and 0.90 for 10 and 30 keV); a $\pm10\%$ change in photoabsorption changed the information by $-4\%$ and $+6\%$, and 2\% energy resolution by $+2\%$. The model's total attenuation coefficient was 6\% below the tabulated value at 511 keV and 5--13\% below it between 100 and 600 keV, mainly because coherent scattering was omitted; raising photoabsorption by 25\%, more than this deficit, increased the information per pair of interacting photons by 10\%, which bounds the effect of the deficit.

The independent GATE simulation confirmed both models. For $2.3\times10^5$ annihilation pairs in which both photons first scattered by Compton scattering, the fitted coefficient was $\kappa=1.039\pm0.016$ with entanglement enabled and $0.536\pm0.017$ with it disabled, reproducing the values 1 and 1/2 of the maximally entangled and the separable orthogonally polarised states within 4\%; the common excess reflects the slightly larger azimuthal modulation of the Livermore polarised Compton model relative to the free-electron Klein--Nishina analysing power used here. For $5\times10^5$ photons incident on a module, GATE gave the same usable fraction as our model (0.266 versus 0.265 with sequence reconstruction; 0.181 versus 0.188 for two clusters), but lower correct-ordering rates (0.65 versus 0.70 for two clusters; 0.74 versus 0.87 for the first pair of multi-cluster events) and a lower information per usable pair (0.0100 versus 0.0126). The difference arises from electron ranges, fluorescence and bremsstrahlung, which displace cluster centroids and add spurious clusters. Relative to the idealised model, the information per pair of interacting photons in GATE was 1.03 with sequence reconstruction and 0.58 with two clusters only; these factors were used for all ring results.

\subsection{In-body pre-scattering}

Without an energy window, 3--6\% of accepted pairs contained a photon scattered in the water cylinder (table~\ref{tab:body}). The CZT photopeak window reduced this to 0.23--0.40\%. The worst-case bias $f_s\kappa$ was then at most 0.0037, and its variation between source positions at most 0.0004, below 10\% of the smallest tissue contrast considered. Without an energy window, the worst-case bias reached 0.052 and varied by up to 0.013 between source positions in the 1.5-cm cylinder, which exceeds the muscle--liver contrast of 0.008.

\subsection{Feasibility in a small-animal setting}

\begin{figure}[h]
\centering\includegraphics[width=0.7\textwidth]{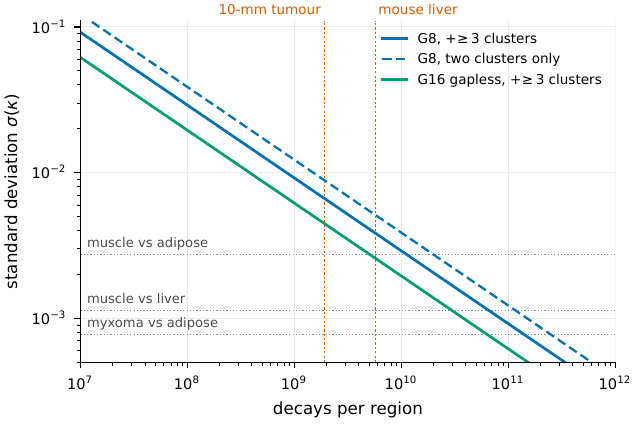}
\caption{Standard deviation of $\kappa$ versus decays per region for the eight-module and gapless 16-module rings (GATE-corrected detector response with sequence reconstruction; dashed, two clusters only). Horizontal dotted lines: precision per region required for a 5$\sigma$ separation of tissue pairs (H0). Vertical lines: all decays of $^{18}$F in a 10-mm tumour and a mouse liver at 400 kBq mL$^{-1}$.}\label{fig:feas}
\end{figure}

Figure~\ref{fig:feas} and table~\ref{tab:req} combine these results, using the GATE-corrected detector response with sequence reconstruction. With G8 the Fisher information per decay was $1.18\times10^{-5}$. Then $\delta\kappa=0.01$ required $8.5\times10^8$ decays per region, about half of the decays available in a 10-mm tumour; $\delta\kappa=0.005$ required $3.4\times10^9$, 1.8 times the available decays. Separating muscle from adipose at 5$\sigma$ under H0 needed $1.1\times10^{10}$ decays per region (6.0 times the tumour's decays, 2.0 times the liver's), muscle from liver $6.6\times10^{10}$ (35 and 12 times), and myxoma from adipose $1.4\times10^{11}$ (74 and 25 times). With two-cluster events only, all requirements increased by a factor of 1.8. The gapless ring reduced all requirements by a factor of 2.2; under H$\eta$ they increased by a factor of 4.

\section{Discussion}

\paragraph{What is measured.} Double-Compton polarimetry of annihilation photons measures one number, the linear-polarisation correlation $\kappa$. Its relation to entanglement requires assumptions about the unmeasured circular correlation: for Bell-diagonal states, $\kappa>1/2$ implies entanglement and bounds the concurrence below by $2\kappa-1$, but the same $\kappa$ is compatible with concurrences up to $\kappa$ (Appendix~A). This is consistent with recent analyses showing that separable states can reproduce the double-Compton cross section and that single-scattering observables admit local hidden-variable models \citep{Hiesmayr2019,Clarke2026,Sutter2026}. For imaging, the distinction is secondary: what carries tissue information is the mixture of channels in (\ref{eq:kobs}), and $\kappa$ is a well-defined, model-independent observable of that mixture. We therefore recommend reporting $\kappa$ (or $R$ together with the analysing power) and reserving the term entanglement degree for statements supported by the bounds above.

\paragraph{Statistical requirements.} The information per double-Compton event is small for reasons that cannot be engineered away: Compton scattering does not resolve circular polarisation, the analysing power at 511 keV does not exceed 0.69, and most scatterings are forward. In the small-modulation limit, (\ref{eq:cfi}) gives $\sigma(\kappa)\approx\sqrt{2/(N\langle(\beta_1\beta_2)^2\rangle)}$, the two-photon analogue of the per-event Stokes uncertainty $\sqrt{2/(\mu^2N)}$ of X-ray polarimetry \citep{Kislat2015} with $\mu^2$ replaced by the mean squared product of analysing powers. Two practical consequences follow. First, scattering-angle windows should not be imposed for estimation; every event should enter the likelihood with its own analysing power. Second, the $R$ ratio, although convenient for display, should not be the estimator: it needed up to 2.3 times as many events as ML in our simulations, because a single mean analysing power is applied to events with widely different $\beta$. The ML and $\beta$-weighted estimators are simple to implement and reached the Cram\'er--Rao bound.

\paragraph{Detector design.} The usable fraction in CZT was limited not by energy or position resolution, which were more than adequate, but by containment of the laterally scattered photon. Tighter packing helped: a gapless ring more than doubled the information per decay, mostly through geometric coverage. Once ordering and position errors are included, two further properties matter: sequence reconstruction of events with three or more clusters, which nearly doubled the information per pair of interacting photons, and the use of the known incident direction for ordering, which gained 19\% over ordering by deposit size. Doppler broadening and energy resolution between 1\% and 2\% were unimportant. Thicker crystals (20 mm: +21\%), finer pixels (1 mm: +12\%) and layered geometries in which the scattered photon is absorbed in a second detector, as in Compton cameras, add further gains. Table~\ref{tab:design} gives the resulting design requirements. Rates of random coincidences were negligible at 10 MBq, and polarisation-correlated Compton events have been measured to have a higher signal-to-random ratio than photoelectric events at clinical activities \citep{Kozuljevic2026}.

\paragraph{Energy resolution and the decoherence question.} Whether a photon keeps its polarisation correlation after scattering in the body is disputed. One view holds that scattering below about 40$^\circ$ does not reduce the correlation \citep{Moskal2026FIH,Ivashkin2023}; another measurement reported a reduction \citep{Watts2021}. With the energy resolution of CZT, pre-scattered photons were reduced to below 0.4\% of accepted pairs, and the worst-case bias varied by less than $4\times10^{-4}$ across positions, so tissue contrasts can be measured whichever view is correct. Without energy discrimination, the worst case varied by more than the muscle--liver contrast across a mouse-sized object, and the variation grows with body size. Energy resolution therefore makes $\kappa$ imaging robust to an open question in the physics of the measurement.

\paragraph{The pick-off question.} Equation~(\ref{eq:kobs}) makes explicit that $\kappa$ carries tissue information only through the pick-off channel, and only when $\kpick<1$. The interpretation of the J-PET result assumes separable pick-off photons \citep{Moskal2025}. Our inversion of their published ratios gives $\kpick$ between 0.34 and 0.58, which excludes maximally entangled pick-off photons and places tissue contrasts at 0.8--1.3 times the values computed under H0. The 3$\sigma$ difference between the two angles shows that a precise value needs the uncorrected counts. A first-principles calculation of $\kpick$ for annihilation with bound molecular electrons is the key theoretical input for the whole programme. Our results show what is at stake: the floors on oxygen estimation are unchanged for any $\kpick<1$, but the required statistics scale as $(1-\kpick)^{-2}$, and the biomarker vanishes at $\kpick=1$. Measurements in materials with known pick-off fractions and different electronic structure, analysed with the estimators recommended here, will discriminate between these hypotheses.

\paragraph{Feasibility.} With $^{18}$F and an eight-module CZT ring, $\kappa$ can be measured to about $\pm0.01$ in a 10-mm tumour, which is sufficient to distinguish separable from maximally entangled pick-off photons in phantoms with large pick-off fractions. It is not sufficient for tissue classification by $\kappa$ alone, which needs 6--74 times the available decays. Classification in such systems will rely on lifetime and three-photon observables, as in \citet{ZhouPMB}, with $\kappa$ contributing as a complementary observable in combined estimation. For that study the present results leave the identifiability floors unchanged but increase the number of decays required in the CZT scenario: by factors of 1.1--1.3 for myxoma versus adipose, where lifetime and three-photon events dominate, by 3--7 for muscle versus liver, and by 5--11 when $\kappa$ is used alone.

\paragraph{Scope of the model.} Coherent scattering, fluorescence, electron transport and polarised Compton kinematics, which our transport model omits, are included in the GATE simulation, and the ring results use the GATE response. Ordering, sequence reconstruction and electron transport enter the rings as a common single-module factor. The effective analysing power is calibrated in reconstructed angle, which in an experiment is done with a calibration source or a validated simulation. The comparison with published data uses reported ratios. The mouse is a uniform water cylinder and the ring source a central point. The channel shares follow the source model of \citet{ZhouPMB} and ex vivo positron-lifetime data, and the post-scattering coefficient is bounded rather than modelled. Detector timing and count-rate parameters not supplied by the manufacturer were taken from \citet{ZhouPMB} and will be measured on the DT-PA05 module.

\section{Conclusion}

Compton polarimetry of annihilation photons extracts less than 1\% of the quantum Fisher information about their polarisation correlation, and a double-Compton event carries about 0.015 units of information on $\kappa$. Using all scattering angles with event-wise maximum likelihood is optimal; angular windows and the $R$ ratio waste up to half the data. The model reproduces published modulation ratios of CZT and scintillator polarimeters and agrees with an independent Geant4 simulation. In a pixelated CZT ring, lateral containment limits the usable events, ordering errors, position errors and electron transport halve the information per event, and sequence reconstruction of multi-site events recovers this loss; energy resolution makes the measurement robust to in-body scattering. With $^{18}$F in a mouse, $\kappa$ can be measured to $\pm0.01$ in centimetre-sized regions, but tissue classification by $\kappa$ alone is out of reach. The biomarker value of $\kappa$ is set by the polarisation state of pick-off photons; published porous-polymer data place $\kpick$ at 0.34--0.58, and a first-principles calculation of $\kpick$ is the next target of theory.

\section*{Acknowledgments}
This work was supported by the National Natural Science Foundation of China (Nos. 62005150, U22A6008 and 82027804), the Shanxi Province Higher Education ``Billion Project'' Science and Technology Guidance Project, the Fundamental Research Program of Shanxi Province (No. 202503021211260), the Natural Science Foundation for Young Scientists of Shanxi Province (No. 202303021222344) and the First Hospital of Shanxi Medical University Hospital Fund Project (No. SYYYRC-2023004).

\section*{Data availability statement}
The code, simulation scripts and numerical results that support the findings of this study are deposited in Zenodo at \href{https://doi.org/10.5281/zenodo.24302646}{https://doi.org/10.5281/zenodo.24302646}. During peer review the files are accessible through a reviewer link provided to the editor; they will be made openly available on publication.

\section*{Author contributions}
Haitao Zhou: conceptualization, methodology, software, formal analysis, validation, visualization, funding acquisition, writing -- original draft, writing -- review and editing. Ruonan Wang: data curation (tissue parameter set), funding acquisition, writing -- review and editing. Jie An, Min Li, Junping Lv, Yaqiong Zhang and Hongliang Wang: writing -- review and editing. Zhifang Wu: resources, writing -- review and editing. Sijin Li: conceptualization, supervision, project administration, funding acquisition, writing -- review and editing.

\section*{Declaration of generative AI use}
During the preparation of this work the authors used Claude (Anthropic) to assist in writing and debugging the Python code for the Fisher-information calculations, the toy Monte Carlo and the detector simulations; to help organise and summarise the literature; and to draft and edit the English text. All assumptions, parameter choices, results and conclusions were designed, checked and verified by the authors, and all cited references were read and verified by the authors. The authors take full responsibility for the content of this article.

\section*{Ethics statement}
This study is entirely theoretical and computational. No human participants, human samples, animals or patient data were involved; tissue parameters were taken from published measurements.

\section*{Conflict of interest}
The authors declare no competing interests.

\appendix
\section{Identifiability: from $\kappa$ to concurrence}\label{app:conc}

\begin{figure}[h]
\centering\includegraphics[width=0.55\textwidth]{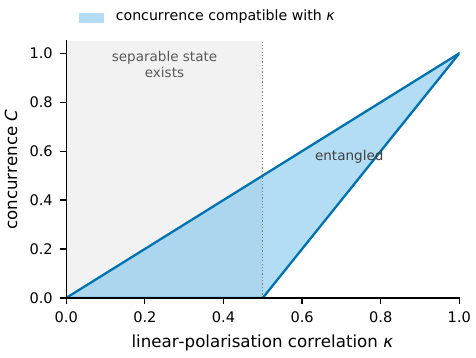}
\caption{Concurrence compatible with a measured $\kappa$ for Bell-diagonal states (Appendix~A).}\label{fig:conc}
\end{figure}

Write the two-photon polarisation state in the Pauli basis as $\rho=\frac14(I+\sum_i r_i\sigma_i\otimes I+\sum_j s_j I\otimes\sigma_j+\sum_{ij}T_{ij}\sigma_i\otimes\sigma_j)$, with $\sigma_{1,2}$ the linear Stokes components and $\sigma_3$ the circular one. We assume no single-photon polarisation ($r=s=0$) and a correlation matrix that is diagonal in a frame in which the annihilation state has $T={\rm diag}(-1,-1,-1)$, i.e.\ a Bell-diagonal state. Rotational invariance about the line of flight requires $T_{11}=T_{22}\equiv-\kappa$, and the double-Compton distribution (\ref{eq:joint}) depends on $T$ only through $\kappa$; the circular correlation $t_3=T_{33}$ is not measured. The Bell weights are $p_0=(1+2\kappa-t_3)/4$ for the annihilation state, $p_{1,2}=(1+t_3)/4$ and $p_3=(1-2\kappa-t_3)/4$. Positivity requires $-1\le t_3\le1-2\kappa$. The concurrence of a Bell-diagonal state is $C=\max(0,2p_{\max}-1)$, which over the allowed range of $t_3$ gives
\begin{equation}
\max(0,\,2\kappa-1)\;\le\;C\;\le\;\kappa,\qquad \kappa\le F_0=p_0\le\frac{1+\kappa}{2},
\end{equation}
where $F_0$ is the fidelity with the annihilation Bell state (figure~\ref{fig:conc}). Thus $\kappa>1/2$ certifies entanglement within this family, and $\kappa=1$ fixes the maximally entangled state, in agreement with \citet{Joos2026}; for $\kappa\le1/2$ a separable state with the same $\kappa$ exists. The family A states of section~\ref{sec:fisher}, mixtures of the annihilation Bell state with the separable state ($t_3=0$ at $\kappa=1/2$), have $t_3=1-2\kappa$; they saturate the positivity bound and therefore carry the minimum concurrence compatible with $\kappa$, $C=2\kappa-1$. Family B has $t_3=-\kappa$.

\section{Fisher information}\label{app:fisher}

For $p(x)=(1-a\cos x)/2\pi$ on $[0,2\pi)$ with $x=2\Delta\varphi$ and $a=\kappa\beta_1\beta_2$, $\partial_\kappa p=-\beta_1\beta_2\cos x/2\pi$ and
\begin{equation}
I_\kappa=(\beta_1\beta_2)^2\frac{1}{2\pi}\int_0^{2\pi}\frac{\cos^2x}{1-a\cos x}\,dx=(\beta_1\beta_2)^2\,\frac{1}{a^2}\left(\frac{1}{\sqrt{1-a^2}}-1\right),
\end{equation}
using $\frac{1}{2\pi}\int_0^{2\pi}(1-a\cos x)^{-1}dx=(1-a^2)^{-1/2}$. For the quantum Fisher information, the states of families A and B are diagonal in the Bell basis with parameter-independent eigenvectors, so $F_Q=\sum_k(\partial_\kappa p_k)^2/p_k$. Family A has weights $(\kappa,\,(1-\kappa)/2,\,(1-\kappa)/2,\,0)$, giving $F_Q=1/[\kappa(1-\kappa)]$; family B has $((1+3\kappa)/4,(1-\kappa)/4,(1-\kappa)/4,(1-\kappa)/4)$, giving (\ref{eq:qfi}).

\clearpage
\section*{Tables}

\begin{table}[h]
\centering\small
\caption{Comparison with published measurements of the modulation ratio $R$ and inferred coefficients.}\label{tab:valid}
\resizebox{\textwidth}{!}{\begin{tabular}{llccc}
\toprule
Measurement & Selection & Model, $\kappa=1$ & Model, $\kappa=1/2$ & Measured\\
\midrule
\citet{Watts2021}, CZT & 70--110$^\circ$ & 1.91 & 1.37 & $1.85\pm0.04$\\
 & 93--103$^\circ$ & 1.98 & 1.40 & $1.95\pm0.07$\\
\citet{Ivashkin2023}, scintillators & 80--100$^\circ$, 22.5$^\circ$ counters & 2.46 & -- & $2.435\pm0.018$\\
\midrule
 & & \multicolumn{2}{c}{Separable reference $\kappa$ (expected 0.5)} & Inferred $\kobs$ / $\kpick$\\
\citet{Moskal2025}, XAD-4 & 82$^\circ$ & \multicolumn{2}{c}{0.508} & $0.796\pm0.016$ / $0.34\pm0.05$\\
 & 94$^\circ$ & \multicolumn{2}{c}{0.512} & $0.869\pm0.019$ / $0.58\pm0.06$\\
\bottomrule
\end{tabular}}
\end{table}

\begin{table}[h]
\centering\small
\caption{Pick-off share of two-photon annihilations $\alpha$ and observed correlation coefficient $\kobs$ at pO$_2$ = 30 mmHg under three hypotheses for the pick-off coefficient.}\label{tab:tissue}
\resizebox{\textwidth}{!}{\begin{tabular}{lcccc}
\toprule
Tissue (source of lifetime data) & $\alpha$ & H0 ($\kpick=1/2$) & H$\eta$ ($\kpick=3/4$) & H1 ($\kpick=1$)\\
\midrule
Human adipose \citep{Moskal2021} & 0.217 & 0.891 & 0.946 & 1\\
Human cardiac myxoma \citep{Moskal2021} & 0.228 & 0.886 & 0.943 & 1\\
Human adipose \citep{Moskal2023} & 0.292 & 0.854 & 0.927 & 1\\
Human cardiac myxoma \citep{Moskal2023} & 0.189 & 0.905 & 0.953 & 1\\
Porcine adipose \citep{Avachat2024} & 0.187 & 0.907 & 0.953 & 1\\
Porcine liver \citep{Avachat2024} & 0.132 & 0.934 & 0.967 & 1\\
Porcine muscle \citep{Avachat2024} & 0.148 & 0.926 & 0.963 & 1\\
\bottomrule
\end{tabular}}
\end{table}

\begin{table}[h]
\centering\small
\caption{Floors on $\sigma({\rm pO}_2)$ (mmHg; $10^{16}$ decays, microstructure nuisances with inter-tissue priors, long-axial-field-of-view scenario of \citet{ZhouPMB}) under the three pick-off hypotheses. L: lifetime events; C: double-Compton events; R: three-photon events.}\label{tab:floor}
\begin{tabular}{lcccccc}
\toprule
 & \multicolumn{3}{c}{Muscle} & \multicolumn{3}{c}{Adipose}\\
\cmidrule(lr){2-4}\cmidrule(lr){5-7}
Hypothesis & L & L+C & L+R+C & L & L+C & L+R+C\\
\midrule
H0 & 2214 & 100 & 8.8 & 339 & 14 & 2.7\\
H$\eta$ & 2214 & 100 & 8.8 & 339 & 14 & 2.7\\
H1 & 2214 & 2214 & 1661 & 339 & 339 & 275\\
\bottomrule
\end{tabular}
\end{table}

\begin{table}[h]
\centering\small
\caption{Usable polarimetric events in DT-PA05 CZT ($E_{\rm th}$ = 15 keV). Single-module fractions are relative to photons that interact in the crystal; ring quantities are per decay. Information per decay includes a live fraction of 0.596 and $b$ = 0.05.}\label{tab:czt}
\resizebox{\textwidth}{!}{\begin{tabular}{lccc}
\toprule
Quantity & Single module & G8 ring & G16 gapless ring\\
\midrule
Interaction probability (photon / both photons) & 0.470 & $2.56\times10^{-2}$ & $4.64\times10^{-2}$\\
Clean two-cluster fraction, all angles & 0.173 & 0.180 & --\\
Clean two-cluster fraction, 82$\pm$20$^\circ$ & 0.066 & 0.067 & --\\
Usable pairs per decay, all angles & -- & $8.7\times10^{-4}$ & $1.8\times10^{-3}$\\
Cross-module share of usable events & -- & 1.7\% & 5.1\%\\
Information per usable pair, all angles & 0.023 & 0.022 & 0.024\\
Fisher information per decay, all angles (ML) & -- & $1.15\times10^{-5}$ & $2.55\times10^{-5}$\\
Fisher information per decay, 82$\pm$20$^\circ$ & -- & $5.8\times10^{-6}$ & --\\
Random coincidences at 10 MBq & -- & 0.6\% & 0.5\%\\
\bottomrule
\end{tabular}}
\end{table}

\begin{table}[h]
\centering\small
\caption{Single DT-PA05 module: idealised and realistic detector models ($E_{\rm th}$ = 15 keV). Fractions are relative to photons that interact in the crystal. Information per pair of interacting photons is the product of the squared usable fraction and the information per usable pair.}\label{tab:real}
\begin{tabular}{p{5.0cm}cccc}
\toprule
Model & Usable & Correct & Info per & Relative\\
 & fraction & order & usable pair & info$^b$\\
\midrule
Idealised (correct order, two clusters) & 0.173 & 1 & 0.023 & 1\\
Realistic, no Doppler, kinematic order, two clusters & 0.189 & 0.70 & 0.013 & 0.68\\
Realistic, Doppler 1.5 a.u., larger-deposit order, two clusters & 0.146 & 0.66 & 0.018 & 0.55\\
Realistic, Doppler 1.5 a.u., kinematic order, two clusters & 0.190 & 0.70 & 0.013 & 0.69\\
Realistic, Doppler 1.5 a.u., kinematic order, $\leq4$ clusters & 0.266 & 0.70 / 0.87$^a$ & 0.013 & 1.26\\
GATE, two clusters & 0.181 & 0.65 & 0.012 & 0.58\\
GATE, $\leq4$ clusters & 0.266 & 0.65 / 0.74$^a$ & 0.010 & 1.03\\
\bottomrule
\end{tabular}

\raggedright\footnotesize $^a$ Two-cluster events / first pair of events with three or four clusters. $^b$ Per pair of interacting photons, relative to the idealised model.
\end{table}

\begin{table}[h]
\centering\small
\caption{Sensitivity of the information per pair of interacting photons to detector parameters (realistic model with sequence reconstruction; reference: 15 mm, 2-mm pixels, 15 keV, 1\% at 662 keV, Doppler 1.5 a.u.).}\label{tab:design}
\begin{tabular}{lcc}
\toprule
Parameter & Value & Relative information\\
\midrule
Crystal thickness & 10 / 20 mm & 0.75 / 1.21\\
Pixel size & 1 / 3 mm & 1.12 / 0.87\\
Energy threshold & 10 / 30 keV & 1.03 / 0.90\\
Photoabsorption cross section & $-10\%$ / $+10\%$ & 0.96 / 1.06\\
Energy resolution at 662 keV & 2\% & 1.02\\
Doppler width & 1 / 2 a.u. & 1.00 / 0.97\\
Cluster ordering & larger deposit instead of kinematic & 0.81\\
Sequence reconstruction & two clusters only & 0.55\\
\bottomrule
\end{tabular}
\end{table}

\begin{table}[h]
\centering\small
\caption{In-body pre-scattering in a water cylinder. $f_s$: fraction of accepted pairs with at least one pre-scattered photon. Differences are relative to an on-axis source in the same cylinder.}\label{tab:body}
\begin{tabular}{lcccccc}
\toprule
 & \multicolumn{3}{c}{CZT photopeak window ($\pm2\sigma$)} & \multicolumn{3}{c}{No energy window}\\
\cmidrule(lr){2-4}\cmidrule(lr){5-7}
Radius / offset (cm) & $f_s$ & $f_s\kappa$ & difference & $f_s$ & $f_s\kappa$ & difference\\
\midrule
1.0 / 0 & 0.0023 & 0.0022 & -- & 0.027 & 0.025 & --\\
1.0 / 0.5 & 0.0025 & 0.0023 & 0.0001 & 0.030 & 0.028 & 0.003\\
1.5 / 0 & 0.0036 & 0.0034 & -- & 0.042 & 0.039 & --\\
1.5 / 0.5 & 0.0040 & 0.0037 & 0.0004 & 0.049 & 0.046 & 0.007\\
1.5 / 1.0 & 0.0038 & 0.0035 & 0.0002 & 0.056 & 0.052 & 0.013\\
\bottomrule
\end{tabular}
\end{table}

\begin{table}[h]
\centering\small
\caption{Decays required per region with $^{18}$F (GATE-corrected detector response with sequence reconstruction; in parentheses, two-cluster events only), and their ratio to all decays in a 10-mm tumour ($1.9\times10^9$) and a mouse liver ($5.7\times10^9$). Tissue separations are 5$\sigma$ under H0 unless stated.}\label{tab:req}
\resizebox{\textwidth}{!}{\begin{tabular}{lcccc}
\toprule
Target & Geometry & Decays per region & Tumour ratio & Liver ratio\\
\midrule
$\delta\kappa=0.01$ & G8 & $8.5\times10^8$ ($1.5\times10^9$) & 0.44 & 0.15\\
 & G16 & $3.8\times10^8$ ($6.8\times10^8$) & 0.20 & 0.07\\
$\delta\kappa=0.005$ & G8 & $3.4\times10^9$ ($6.0\times10^9$) & 1.8 & 0.59\\
 & G16 & $1.5\times10^9$ ($2.7\times10^9$) & 0.80 & 0.27\\
Muscle vs adipose ($\Delta\kappa=0.019$) & G8 & $1.1\times10^{10}$ ($2.0\times10^{10}$) & 6.0 & 2.0\\
 & G16 & $5.1\times10^9$ ($9.1\times10^9$) & 2.7 & 0.90\\
Muscle vs liver ($\Delta\kappa=0.008$) & G8 & $6.6\times10^{10}$ ($1.2\times10^{11}$) & 35 & 12\\
 & G16 & $3.0\times10^{10}$ ($5.3\times10^{10}$) & 16 & 5.2\\
Myxoma vs adipose ($\Delta\kappa=0.0055$) & G8 & $1.4\times10^{11}$ ($2.5\times10^{11}$) & 74 & 25\\
 & G16 & $6.3\times10^{10}$ ($1.1\times10^{11}$) & 33 & 11\\
Muscle vs liver, H$\eta$ ($\Delta\kappa=0.004$) & G8 & $2.6\times10^{11}$ ($4.7\times10^{11}$) & 139 & 46\\
\bottomrule
\end{tabular}}
\end{table}

\clearpage
\section*{Figures}

\begin{figure}[h]
\centering\includegraphics[width=\textwidth]{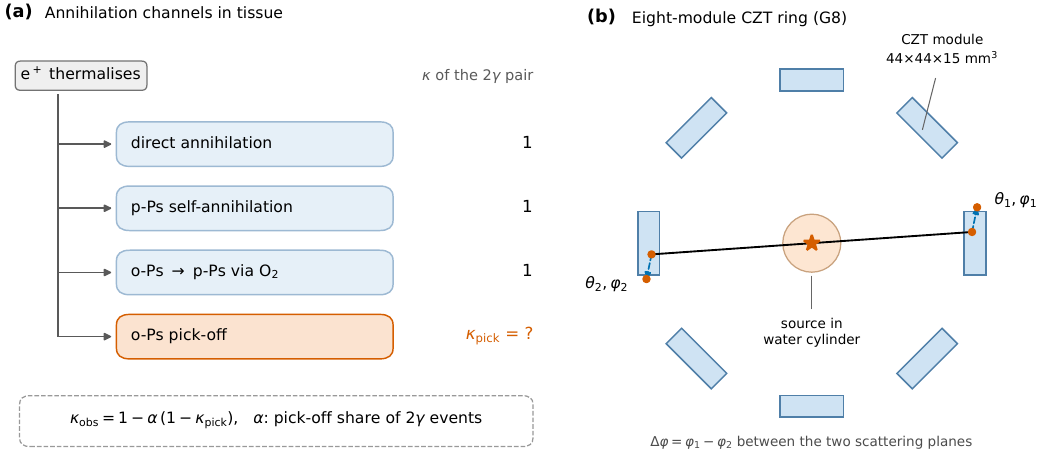}
\caption{(a) Annihilation channels and their polarisation-correlation coefficients; the pick-off coefficient is unknown. (b) Double-Compton measurement in an eight-module CZT ring with a source in a water cylinder.}\label{fig:schem}
\end{figure}

\end{document}